\documentclass[aps,pra,twocolumn]{revtex4}
\usepackage{graphicx}
\usepackage{dcolumn}
\usepackage{bm}
\usepackage{float}
\usepackage[]{hyperref}
\usepackage{color} 

\begin{document}

\title{Demonstration of spatial-light-modulation-based four-wave mixing in cold atoms}
\author{Jz-Yuan Juo,$^{1}$ Jia-Kang Lin,$^{1}$ Chin-Yao Cheng,$^{1}$ Zi-Yu Liu,$^{1}$  Ite A. Yu,$^{2,3}$ and Yong-Fan Chen$^{1,3,}$}
\email{yfchen@mail.ncku.edu.tw}
\affiliation{$^1$Department of Physics, National Cheng Kung University, Tainan 70101, Taiwan \\
$^2$Department of Physics, National Tsing Hua University, Hsinchu 30013, Taiwan\\
$^3$Center for Quantum Technology, Hsinchu 30013, Taiwan}
\date{\today}
\begin{abstract}
Long-distance quantum optical communications usually require efficient wave-mixing processes to convert the wavelengths of single photons. Many quantum applications based on electromagnetically induced transparency (EIT) have been proposed and demonstrated at the single-photon level, such as quantum memories, all-optical transistors, and  cross-phase modulations. However, EIT-based four-wave mixing (FWM) in a resonant double-$\Lambda$ configuration has a maximum conversion efficiency (CE) of 25\% because of absorptive loss due to spontaneous emission. An improved scheme using spatially modulated intensities of two control fields has been theoretically proposed to overcome this conversion limit. In this study, we first demonstrate wavelength conversion from 780 to 795 nm with a 43\% CE by using this scheme at an optical density (OD) of 19 in cold $^{87}$Rb atoms. According to the theoretical model, the CE in the proposed scheme can further increase to 96\% at an OD of 240 under ideal conditions, thereby attaining an identical CE to that of the previous non resonant double-$\Lambda$ scheme at half the OD. This spatial-light-modulation-based FWM scheme can achieve a near-unity CE, thus providing an easy method of implementing an efficient quantum wavelength converter for all-optical quantum information processing.
\end{abstract}




\maketitle



\newcommand{\FigOne}{
    \begin{figure}[t] 
    \includegraphics[width=8.5cm]{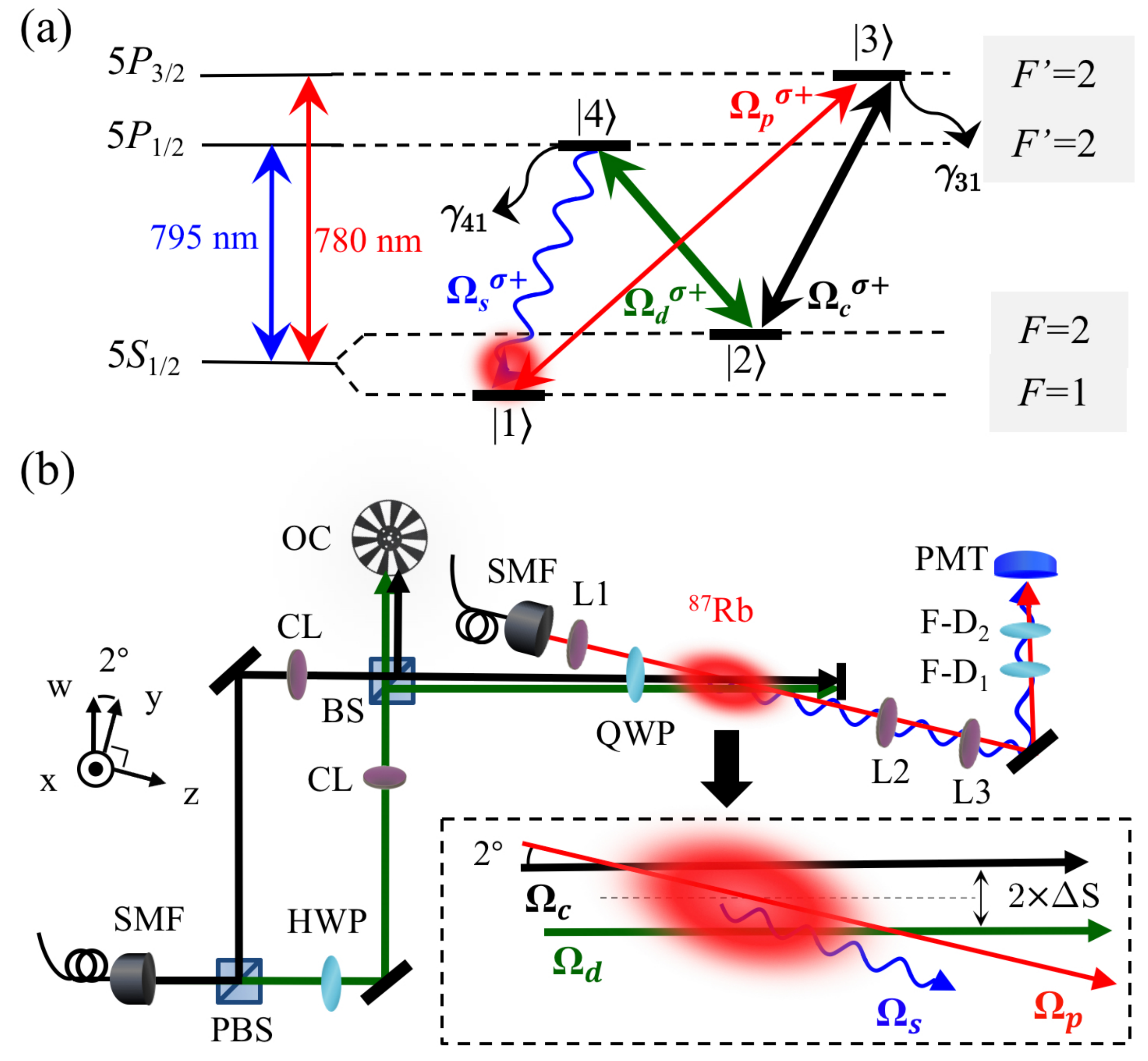}
    \caption{
Energy level diagram and experimental setup. (a)  $^{87}$Rb four-level system in our experiments. All fields are on resonance and selected as right circular polarization ($\sigma^+$). $\Omega_p$ and $\Omega_c$ form the EIT system in the D$_2$-line transition (780 nm). After applying $\Omega_d$, $\Omega_s$ is subsequently generated in the D$_1$-line transition (795 nm), thereby forming the double-$\Lambda$ FWM system. (b) The experimental setup. BS, beam splitter; PBS, polarization beam splitter; QWP, quarter-wave plate; HWP, half-wave plate; SMF, single-mode fiber; L1 and L2, lens with focal length of 400 mm; L3, lens with focal length of 500 mm; CL, cylindrical lens with focal length of 500 mm used for one-dimensional shaping of the light; PMT, photomultiplier tube; F-D$_1$, wavelength filter for $^{87}$Rb D$_1$-line transition; F-D$_2$, wavelength filter for $^{87}$Rb D$_2$-line transition; OC, optical chopper; $\Delta$S, separation distance between the peak position of the control beam and center of the cold atomic cloud.}
    \label{fig:Experimental setup}
    \end{figure}
}


\newcommand{\FigTwo}{
    \begin{figure}[t] 
    \includegraphics[width=8.8cm]{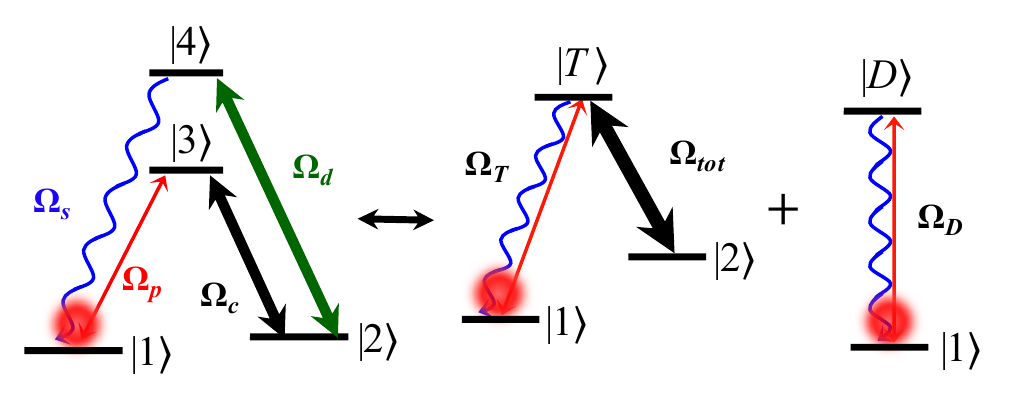}
    \caption{
Transition scheme of double-$\Lambda$-based FWM. $\Omega_p$ and $\Omega_c$ form one $\Lambda$ configuration and $\Omega_s$ and $\Omega_d$ form another. After transforming to normal modes, $\Omega_T$ and $\Omega_D$ represent the physical channels of the transmission mode and dissipation mode in the system, respectively. 
    }
    \label{fig:Theoretical model}
    \end{figure}
}


\newcommand{\FigThree}{
    \begin{figure}[t] 
    \includegraphics[width=8.5cm]{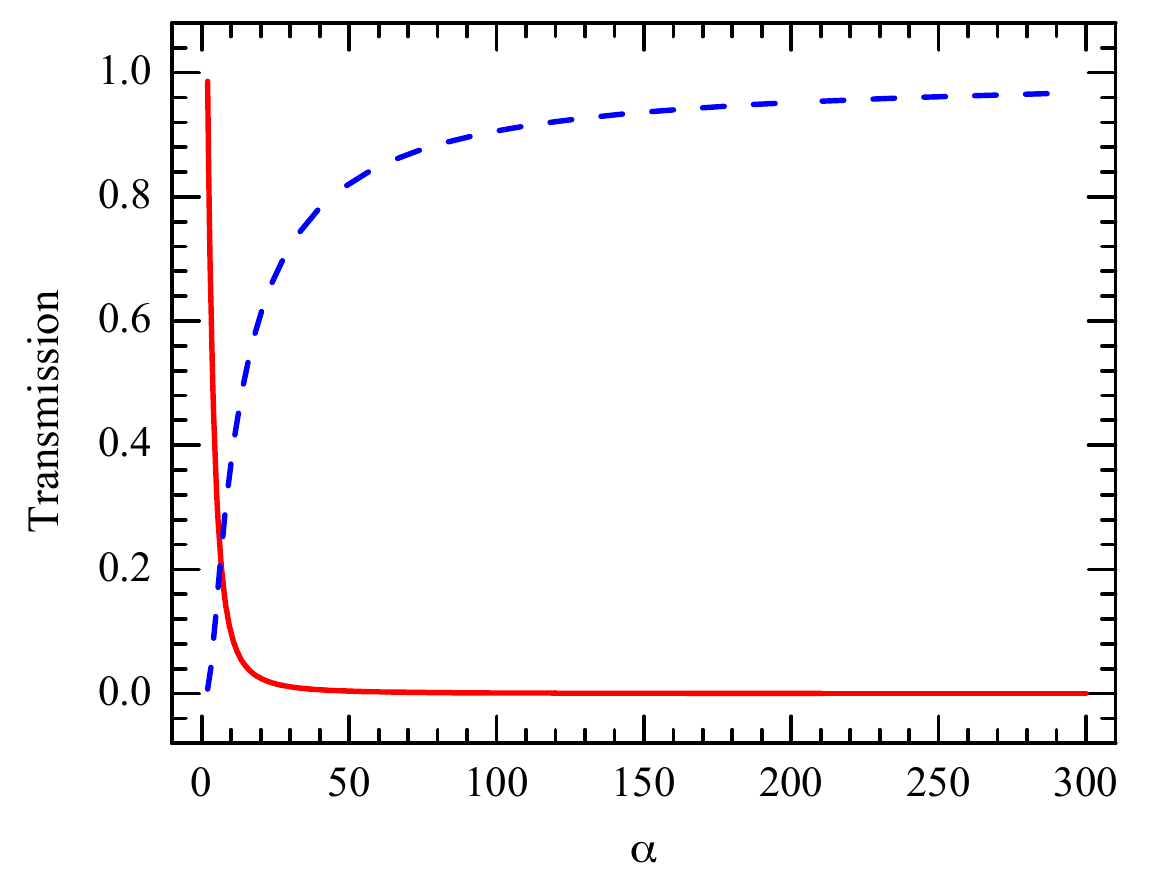}
    \caption{
Transmissions of the probe and signal fields in the SLM-based FWM versus the OD. The red (solid) and blue (dashed) lines are the theoretical curves of Eqs. (13) and (14), respectively.}
    \label{fig:Theoretical curves}
    \end{figure}
}


\newcommand{\FigFour}{
    \begin{figure}[t] 
    \includegraphics[width=8.8cm]{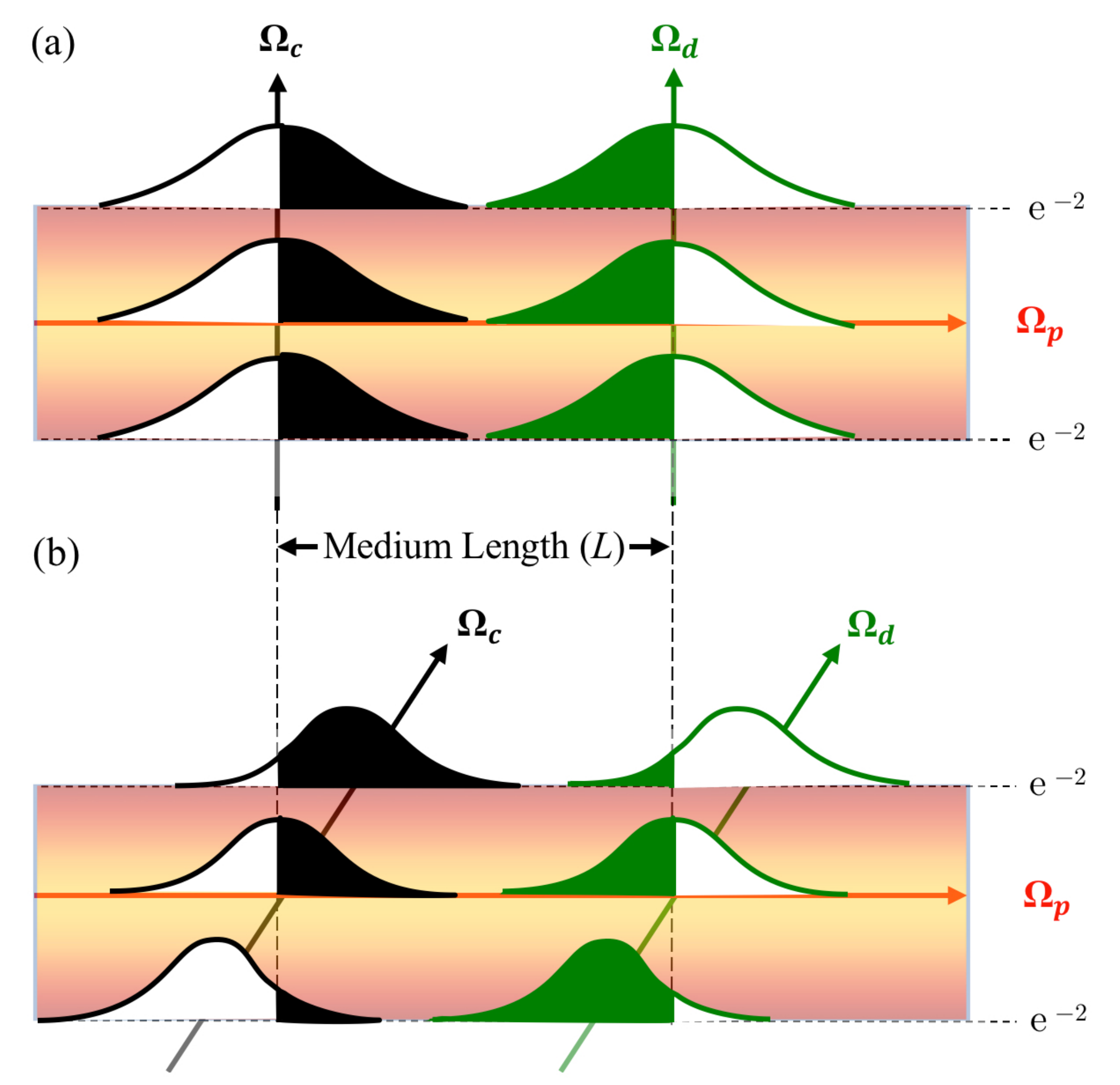}
    \caption{
Intensity-mismatch effect in the SLM-based FWM system. (a) The ideal condition is 90$^\circ$. (b) The experimental condition is narrower than 90$^\circ$. Solid lines with arrows indicate the peak strength and propagating direction of laser beams. The area between dashed lines is the assumptive area of an atomic cloud. If the $e^{-2}$ diameter of $\Omega_p$ is considered, the intensity-mismatch effect is evident on the edge of $\Omega_p$.
    }
    \label{fig:Intensity mismatch}
    \end{figure}
}


\newcommand{\FigFive}{
    \begin{figure}[t] 
    \includegraphics[width=8.5cm]{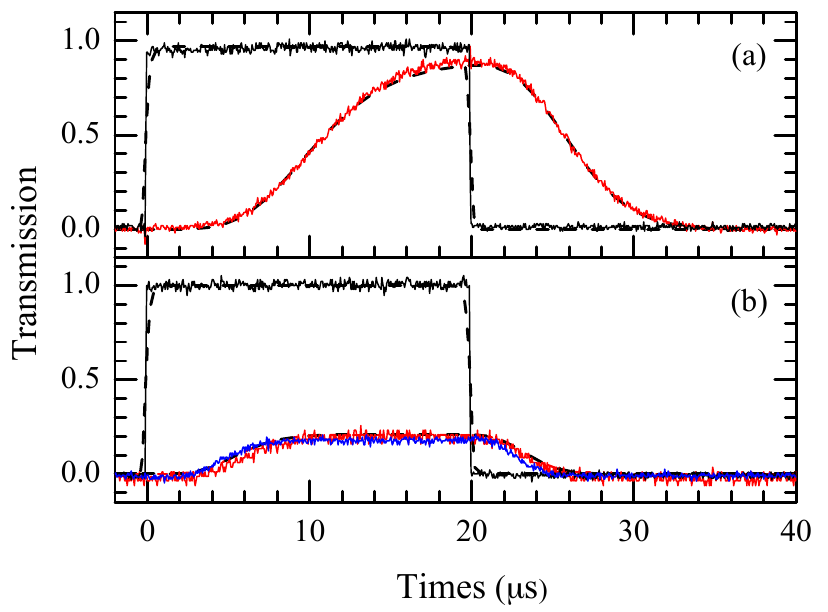}
    \caption{
EIT and FWM observations in the pulsed regime without SLM. The black solid lines are the incident probe pulses. The red and blue solid lines are the transmitted probe and generated signal pulses, respectively. The dashed lines are the theoretical curves with the following parameters: (a) $\alpha= 19$, $\Omega_c= 0.26\Gamma$, $\Omega_d= 0\Gamma$, and $\gamma_{21}=5\times 10^{-4}\Gamma$; (b) $\alpha= 19$, $\Omega_c=\Omega_d= 0.26\Gamma$, and $\gamma_{21}=1.0\times 10^{-3}\Gamma$. The $e^{-2}$ waists (radiuses) of $\Omega_c$ and $\Omega_d$ are approximately 2.1 mm when two CLs are not used. 
    }
    \label{fig:Normal FWM}
    \end{figure}
}


\newcommand{\FigSix}{
    \begin{figure}[t] 
    \includegraphics[width=8.7cm]{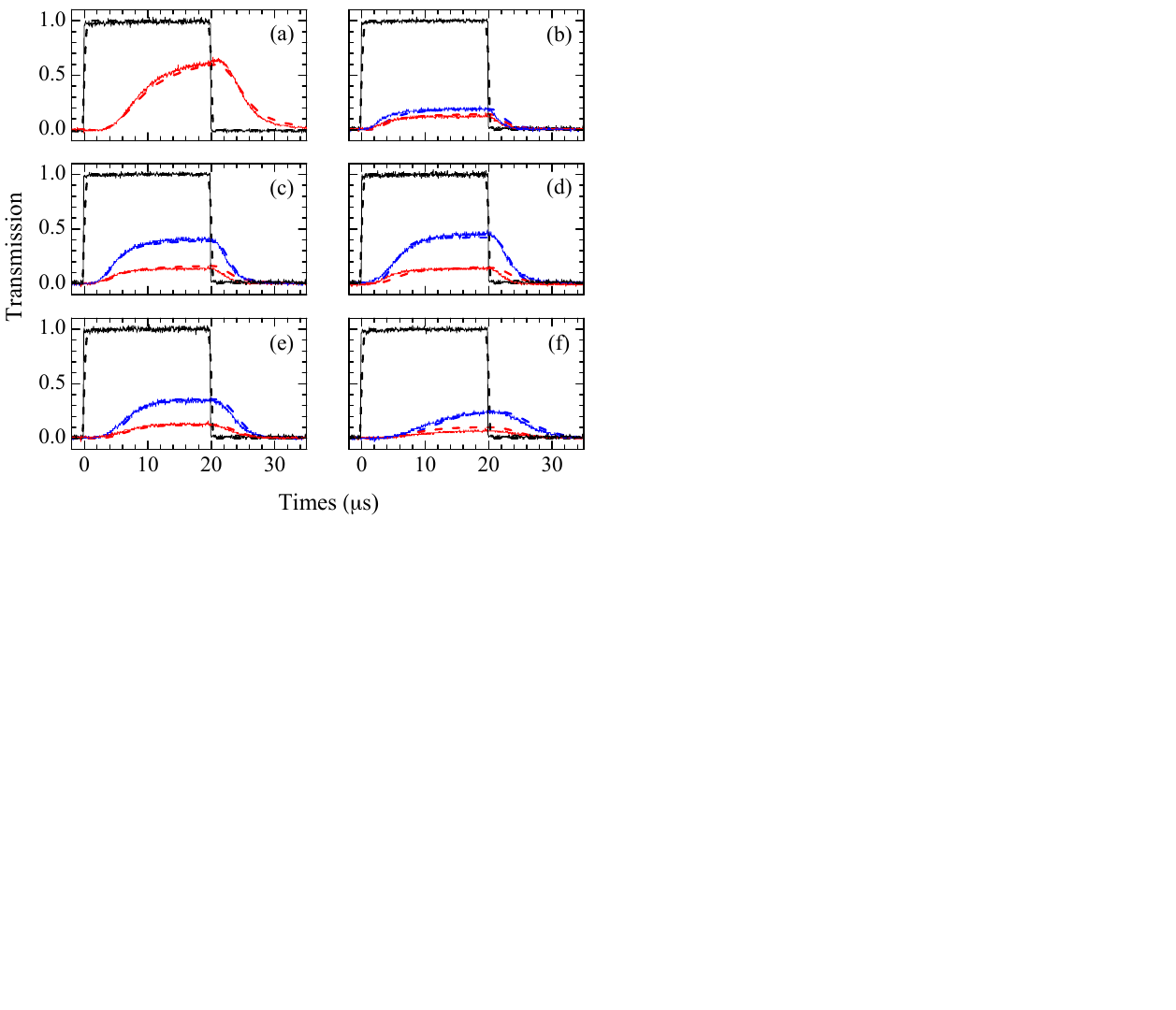}   
    \caption{
EIT and FWM observations in the pulsed regime with SLM. The black solid lines are the incident probe pulses. The red and blue solid lines are the transmitted probe and generated signal pulses, respectively. The dashed lines are the theoretical curves with the following parameters: $\alpha = 19$, $\Omega_c = 0.39\Gamma$; (a) $\Omega_d = 0\Gamma$ and $\gamma_{21} = 4 \times 10^{-4}\Gamma$; (b) $\Delta$S = 3 $\mu$m, $\Omega_d = 0.42\Gamma$, and $\gamma_{21} = 1.0 \times 10^{-3}\Gamma$; (c) $\Delta$S = 30 $\mu$m, $\Omega_d = 0.37\Gamma$, and $\gamma_{21}= 6 \times 10^{-4}\Gamma$; (d) $\Delta$S = 54 $\mu$m, $\Omega_d = 0.38\Gamma$, and $\gamma_{21} = 5 \times 10^{-4}\Gamma$; (e) $\Delta$S = 75 $\mu$m, $\Omega_d = 0.42\Gamma$, and $\gamma_{21} = 8 \times 10^{-4}\Gamma$; (f) $\Delta$S = 95 $\mu$m, $\Omega_d = 0.45\Gamma$, and $\gamma_{21} = 1.0 \times 10^{-3}\Gamma$. Compared with the information in Fig.~\ref{fig:Normal FWM}, here, the $e^{-2}$ waists (radiuses) of $\Omega_c$ and $\Omega_d$ on the w-axis are focused to approximately 124 $\mu$m by two CLs.
}
    \label{fig:SLM FWM}
    \end{figure}
}


\newcommand{\FigSeven}{
    \begin{figure}[H] 
    \includegraphics[width=8.3cm]{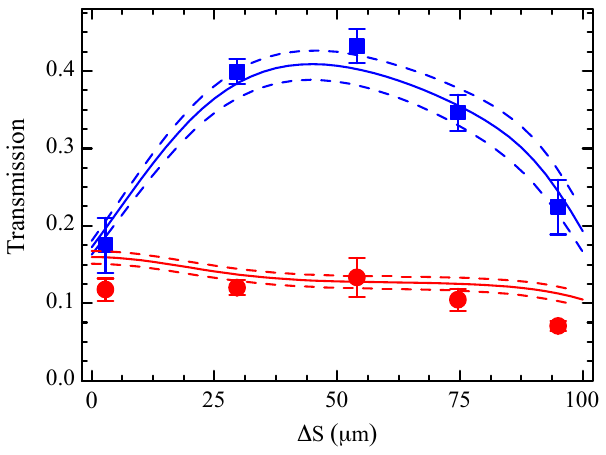}
    \caption{
Transmissions of the probe and signal pulses in the SLM-based FWM as a function of $\Delta$S. The red circles and blue squares are the probe and signal transmissions, respectively. The error bar represents a $\pm$1 standard deviation based on five successive measurements. The solid lines denote theoretical curves with the following parameters: $\alpha = 19$, $\Omega_{c} = 0.39\Gamma$, $\Omega_{d} = 0.41\Gamma$, and $\gamma_{21} = 8 \times 10^{-4}\Gamma$. The dashed lines indicate theoretical curves with the $\pm$1 standard deviation of $\gamma_{21}$[i.e., $\gamma_{21} = 5 \times 10^{-4}\Gamma$ (upper line) and $\gamma_{21} = 1.1 \times 10^{-3}\Gamma$ (lower line)].
}
    \label{fig:SLM FWM Efficiency}
    \end{figure}
}


\section{Introduction}
Photons have been proven to be the optimal qubit carriers to transmit information for quantum communications and building blocks in the field of linear optical computation~\cite{Duan01,Tanzilli05,Azuma15}. Because of the presence of telecommunications optical fibers, wavelengths of photons of approximately 1550 nm exhibit the minimum absorption in fiber-optic quantum information networks~\cite{Radnaev10,Pelc11}. However, although information encoded in photons can be effectively stored, controlled, and processed using quantum nonlinear processes such as electromagnetically induced transparency (EIT), the optical wavelengths in the alkaline atomic interface are mostly 600-800 nm~\cite{HarrisEIT,Liu01,LukinEIT,FleischhauerEIT,Eisaman05,Chen06,Chen13,Hsiao18}. These implementations based on diverse physical systems suffer wavelength incompatibility between components. To overcome such difficulties, an efficient quantum frequency conversion (QFC) device for optical quantum information processing (QIP) is required to serve as the joint for quantum processors that use different physical systems in various wavelength regimes~\cite{Kumar90,Huang92,Pelc10}. 

Because EIT features extremely high dispersion and considerable suppression of linear absorption, various nonlinear optical effects based on EIT have been studied and experimentally demonstrated at the photon level~\cite{Harris99,Zhu03,Braje03,Chen05,LoXPM,Chen12,W. Chen,H. Gorniaczyk,D. Tiarks14,C. Adams,SteinbergXPM,K. M. Beck,D. Tiarks16,Liu16,S. Parkins}. Recently, four-wave mixing (FWM) based on a double-$\Lambda$ configuration has attracted much attention because of its potential to enable implementation of a practical QFC device and other promising applications~\cite{Harris96,Lukin98,Lukin99,Harris4WM,Deng02,Wu04,Lett4WM,Xiao6WM1,Xiao6WM2,Novikova4FM,Chen4WM,Xiao4WM1,Moon4WM}. By using the non resonant double-$\Lambda$ FWM scheme, Wang {\it et al.} obtained a 73\% conversion efficiency (CE) in hot rubidium atoms with continuous-wave lasers of intensities greater than 1 W/cm$^2$~\cite{Wang10}. Chiu {\it et al.} further achieved such frequency conversion at low-light levels of 1.8 mW/cm$^2$ in cold rubidium atoms and observed a 46\% CE, which successfully overcame the conversion limit (25\%) of forward resonant double-$\Lambda$ scheme~\cite{Chiu14}. To preserve the strong interaction coupling between atoms and photons, Kang {\it et al.} demonstrated a 10\% CE in their proposed resonant double-$\Lambda$ scheme by using a so-called backward FWM to reverse the propagating direction of control field~\cite{Kang04}. Because all laser fields are resonant with the atomic transitions, the interaction strength between atoms and photons is as strong as it can be, and a higher CE can be obtained when using a smaller optical density (OD). According to theoretical predictions, the CE in this backward resonant scheme can achieve the same value as that in the previous non resonant scheme with only 50\% of the OD. Recently, Liu {\it et al.} have demonstrated a CE of 63\% at an OD of 48 by using this backward FWM scheme~\cite{Liu17}.

An improved FWM scheme using spatially modulated intensities of two control laser fields in a forward configuration was theoretically proposed by Lee {\it et al.}~\cite{Lee16}. In this scheme, the conversion limit in resonant FWM processes can be exceeded, and the CE increases to 96\% at an OD of 240. In addition, the CE in this forward scheme is not restricted by the phase-mismatch effect because the angles between the laser fields can be arranged for narrow angle conditions. Although phase mismatching can be compensated through adjusting two-photon detuning in the double-$\Lambda$ scheme~\cite{Harris96}, the impact of phase mismatching in the proposed scheme is relatively slight compared with that in the backward scheme. We first demonstrate the observation of wavelength conversion from 780 to 795 nm by using a spatial light modulation (SLM)-based FWM scheme. In our experimental observations, we obtain a CE of 43\% at an OD of 19 that apparently overcomes the conversion limit  in the resonant FWM processes. Furthermore, according to theoretical predictions, a near-unity CE can be achieved when a sufficiently large OD is applied under ideal conditions. Such a highly efficient wavelength conversion scheme based on EIT provides a simple method for constructing a narrowband QFC device for QIP.


\section{Experimental Details} \label{Sec:exp}

In this study, we conduct EIT-based FWM experiments in a cold $^{87}$Rb atomic system by using a typical magneto-optical trap (MOT). The transitions driven by each laser field are depicted in Fig.~\ref{fig:Experimental setup}(a). The entire atomic population is initially optically pumped into the ground state $|1\rangle$. The probe field ($\Omega_p$ represents its Rabi frequency) with a peak intensity of approximately 1 $\mu$W/cm$^2$ drives the transition of the state $|1\rangle\leftrightarrow|3\rangle$ and forms a $\Lambda$-type EIT system along side the coupling field ($\Omega_c$), which drives the transition of the state $|2\rangle\leftrightarrow|3\rangle$ with a peak intensity of approximately 1 mW/cm$^2$. Therefore, most of the population remains in the ground state $|1\rangle$ throughout the experiment. Under the EIT condition, the driving field ($\Omega_d$) with the same peak strength of $\Omega_c$ drives the transition of state $|2\rangle \leftrightarrow |4\rangle$ and induces the FWM process to generate the signal field ($\Omega_s$) during the transition of state $|1\rangle \leftrightarrow |4\rangle$. The wavelength is consequently converted from 780 ($\Omega_p$) to 795 ($\Omega_s$) nm.

\FigOne

The main apparatus used in the experiments is shown in Fig.~\ref{fig:Experimental setup}(b). The probe and coupling beams are produced by two different diode lasers injection locked together by an external cavity diode laser (ECDL) to ensure their phase coherence. However, the driving beam is produced by another ECDL that is not phase coherent with the probe and coupling beams. Before coupling all the laser beams into single-mode fibers (SMFs), we send them through acousto-optic modulators to modulate their optical frequencies and control the timing sequence. After emission from the SMFs, the profiles of the probe, coupling and driving beams are shaped into the Gaussian mode and their optical paths are stabilized.

Subsequently, the coupling and driving beams are split by a polarization beam splitter (PBS) and are subsequently focused on the center of the cold atomic cloud by using two cylindrical lenses (CLs), which are set to focus the waists of the two control beams ($\Omega_c$ and $\Omega_d$) only on the w-axis, as shown in Fig.~\ref{fig:Experimental setup}(b). A half-wave plate (HWP) and quarter-wave plate (QWP) are set specifically to make all the laser fields right-circularly polarized ($\sigma +$). F-D$_1$ and F-D$_2$ are wavelength filters of 795 and 780 nm, respectively, and  distinguish $\Omega_s$ and $\Omega_p$, respectively, when each measurement is executed. The separation distance (2$\Delta$S) between the peak position of the two control beams is controlled by adjusting the position of the aforementioned two CLs mounted at their respective three-dimensional (3D) translation stages. In addition, notably, a small angle of approximately 2$^\circ$ is maintained between the probe and two control beams throughout the experiment. 

To reduce the light leakage from $\Omega_c$ and $\Omega_d$, we use a pinhole positioned after the MOT to allow only $\Omega_p$ and $\Omega_s$ to pass. Finally, a photomultiplier tube (PMT) at the output of the MOT measures the transmitted probe and the generated signal pulses. According to the phase-match condition, $\Omega_s$ is, in principle, generated in the direction of the angle of approximately 0.05$^\circ$ to the $\Omega_p$. To ensure that $\Omega_p$ and $\Omega_s$ are collected in their entirety by the PMT that is positioned approximately 1.5 m from the MOT, we use two lenses (L2 and L3) to refocus $\Omega_p$ and $\Omega_s$. The waists and beam profiles of the coupling and driving fields are measured using an optical chopper (OC) and the rotating knife-edge technique. Notably, the distance of the optical path from the center of the atomic cloud to the OC is the same as the focal length of both CLs. Thus, we can determine the spatial relationship between the two control beams presented in the atomic cloud by observing their positions on the edge of OC.

All figures based on the experimental data in the following section are averages of 1024 times; for each measurement, the pulse repetition rate is 100 Hz. The incident probe, transmitted probe, and generated signal pulses are measured separately throughout the experiment. The atomic medium length ($L$) is approximately 3.5 mm on the z-axis. The measured $e^{-2}$ waists (radiuses) of $\Omega_c$ and $\Omega_d$ are both approximately 2.1 mm on the x-axis and 124 $\pm$ 1 $\mu$m on the w-axis when two CLs are used. In addition, by using L1, $\Omega_p$ is focused on the center of the cold atomic cloud with the $e^{-2}$ waist of 141 $\pm$ 1 $\mu$m.


\section{Theoretical Model} \label{Sec:thy}

The theoretical model of the SLM-based FWM was studied in~\cite{Lee16}. In this section, we simply interpret the physical picture of this novel FWM scheme. First, we consider a medium consisting of double-$\Lambda$ type four-level atoms, as plotted in Fig.~\ref{fig:Theoretical model}. The dynamic behaviors of the probe and signal pulses propagating inside the FWM medium can be described using the Maxwell-Schr\"{o}dinger equations as follows:
\begin{eqnarray}
\frac{\partial\Omega_{p}}{\partial z} + \frac{1}{c}\frac{\partial\Omega_{p}}{\partial t} = i \frac{\alpha_{p}\gamma_{31}}{2L} \rho_{31},\label{Eq.slowP}\\
\frac{\partial\Omega_{s}}{\partial z} + \frac{1}{c}\frac{\partial\Omega_{s}}{\partial t} = i \frac{\alpha_{s}\gamma_{41}}{2L}
\rho_{41},\label{Eq.slowS}
\end{eqnarray}
where $\rho_{ij}$ represents the atomic coherence between $|i\rangle$ and $|j\rangle$; $\gamma_{ij}$ describes the total coherence decay rates from the excited state $|i\rangle$ to state $|j\rangle$;  and $\alpha_{p}$ = $n\sigma_{31}L$ ($\alpha_{s}$ = $n\sigma_{41}L)$ is the OD of the probe (signal) transition, where $n$ is the density of the atoms, $\sigma_{ij}\propto\lambda_{ij}^2$ is the absorption cross section of the atoms, and $\lambda_{ij}$ is the relevant transition wavelength. By considering three degenerate Zeeman sublevels of the ground state $|1\rangle$ in our experiment, the average Clebsch-Gordan coefficients of the probe and signal transitions are equal. Therefore, the ODs of the probe and signal transitions are approximately the same (i.e., $\sigma_{31}\simeq\sigma_{41}$) because the wavelength difference is minor in the rubidium atomic system.

To simplify the model, we consider the condition of $\alpha_{p}=\alpha_{s}=\alpha$ in the following equations. After considering the slowly varying amplitudes of the density matrix elements and the first-order perturbation of the probe and signal fields, the optical Bloch equations for the double-$\Lambda$ FWM system are expressed as
\begin{eqnarray}
\frac{d}{dt}\rho_{41} = \frac{i}{2}\Omega_{s} + \frac{i}{2}\Omega_{d}\rho_{21}-\frac{\gamma_{41}}{2}\rho_{41},\\\label{Eq.p41} \frac{d}{dt}\rho_{31} = \frac{i}{2}\Omega_{p} + \frac{i}{2}\Omega_{c}\rho_{21} -\frac{\gamma_{31}}{2}\rho_{31},\\\label{Eq.p31} \frac{d}{dt}\rho_{21} = \frac{i}{2}\Omega^{\ast}_{c}\rho_{31} +\frac{i}{2}\Omega^{\ast}_{d}\rho_{41} -
\frac{\gamma_{21}}{2}\rho_{21},\label{Eq.p21}
\end{eqnarray}
where $\gamma_{21}$ is the dephasing rate of ground states $|1\rangle$ and $|2\rangle$ and the initial condition of population $\rho_{11}=1$ is assumed. We determine the aforementioned experimental parameters as follows: $\Omega_c$ was obtained by measuring the two absorption peak in the EIT spectrum \cite{XiaoEIT}. $\alpha$ was determined by measuring the delay time of EIT slow light in the pulse regime \cite{Hau99}. $\gamma_{31}=\gamma_{41}\approx 1.25 \Gamma$ was estimated by measuring the spectral width of two-level absorption, which was contributed primarily by the spontaneous emission ($\Gamma \simeq 2\pi\times 6$ MHz) and laser linewidth of approximately 1 MHz in our experiment~\cite{Lo10}. $\Omega_d$ was estimated by numerically fitting the transmitted profiles of the probe and signal pulses propagating inside the FWM medium. $\gamma_{21}$ was adjusted to fit the probe and signal transmissions.

\FigTwo

To understand the physics of the EIT-based double-$\Lambda$ FWM system, we can transform the system by introducing a picture of the normal modes. Such a system contains two modes of light fields, denoted as $\Omega_T $ and $ \Omega_D$ and given by
\begin{eqnarray}
\left[
\begin{array}{clr}
\Omega_{T}\\
\Omega_{D} 
\end{array}
\right]
=
\frac{1}{\Omega_{tot}}
\left[
\begin{array}{clr}
\Omega_{c}^{*} & \Omega_{d}^{*}\\
-\Omega_{d} &    \Omega_{c}
\end{array}
\right]
\left[
\begin{array}{clr}
\Omega_p\\
\Omega_s\\  
\end{array}
\right],
\end{eqnarray}
where $\Omega_{tot}=\sqrt{|\Omega_{c}|^2+|\Omega_{d}|^2}$ denotes the total strength of the two control fields. Likewise, excited states $|3\rangle$ and $|4\rangle$ are transformed into $|T\rangle$ and $|D\rangle$, which are given by
\begin{eqnarray}
|T\rangle = \frac{\Omega_{c}^{*}}{\Omega_{tot}}|3\rangle + \frac{\Omega_{d}^{*}}{\Omega_{tot}}|4\rangle, \\\
|D\rangle = -\frac{\Omega_{d}}{\Omega_{tot}}|3\rangle + \frac{\Omega_{c}}{\Omega_{tot}}|4\rangle.
\end{eqnarray}
After altering the basis, as shown in Fig.~\ref{fig:Theoretical model}, the double-$\Lambda$ system becomes the combination of the two normal modes, namely the transmission mode (TM) and dissipation mode (DM). In the TM system, a $\Lambda$-type EIT formed by $\Omega_T$ and $\Omega_{tot}$ can prevent $\Omega_T$ from being dissipated by vacuum fluctuations. By contrast, in the DM system, a two-level absorption induced by $\Omega_D$ causes the loss of $\Omega_D$ with a magnitude of $e^{-\alpha/2}$. To exemplify, we first consider the condition of the uniform intensities of two control fields on the z-axis, namely $\Omega_c=\Omega_d=\frac{1}{\sqrt{2}}\Omega_{tot}$. By the boundary conditions of $\Omega_p(z=0)=\Omega_{p0}$ and $\Omega_s(z=0)=0$ and solving Eqs. (1)-(5) by assuming that $\gamma_{21}=0$ and $\gamma_{31}=\gamma_{41}$, the steady-state solutions for the fields of the TM and DM are given by
\begin{eqnarray}
\Omega_T(z)=\frac{\Omega_{p0}}{\sqrt{2}},\\
\Omega_D(z)=-\frac{\Omega_{p0}}{\sqrt{2}}e^{-\frac{\alpha}{2L}z}. 
\end{eqnarray}
As shown, half the energy in the incident probe field ($\Omega_{p0}$) is coupled into the DM; in other words, 50\% of total energy is dissipated as the OD increases. Therefore, after reverting to the original basis through Eq. (6), only a 25\% CE is achieved if we define the steady-state CE as the intensity ratio of the generated signal field to the incident probe field; this is the so-called conversion limit in the forward resonant double-$\Lambda$ FWM system \cite{Deng02}.

To surpass this limit to enhance the CE, we must suppress the energy loss caused by the DM (i.e., $\Omega_D$ approaches 0). The SLM-based FWM scheme proposed by Lee {\it et al.} successfully provided a solution. The researchers suggested that the condition of $\Omega_D \approx 0$ can be satisfied by using the intensity balance condition [i.e., $\Omega_p(z) \Omega_d=\Omega_s(z) \Omega_c$] from Eq. (6). $\Omega_p$ decreases in the medium while $\Omega_s$ converted from $\Omega_p$ gradually increases, and thus a condition that modulates the strength of $\Omega_{c}$ decreasing and $\Omega_{d}$ increasing in the propagating direction of $\Omega_p$ must be artificially created. Notably, a similar approach that uses the intensity balance condition to overcome the conversion limit was demonstrated by using two counter-propagating control fields \cite{Liu17}.

To test the aforementioned idea, we subsequently consider the condition of the spatially varying intensities of the two control fields in the FWM system, which are given by
\begin{eqnarray}
\Omega_c(z)=\Omega_0\cos{(\beta z)},\\
\Omega_d(z)=\Omega_0\sin{(\beta z)},  
\end{eqnarray}
where $\beta=\frac{\pi}{2L}$ and $\Omega_0$ is the normalized strength of the two control fields, and thus $\Omega_c(0)=\Omega_d(L)=\Omega_0$ and $\Omega_c(L)=\Omega_d(0)=0$. 
By substituting the modulated $\Omega_{c}$ and $\Omega_{d}$ into Eqs. (1)-(5), the steady-state solutions for the probe and signal fields are given by
\begin{eqnarray}
|\Omega_p(z)|^2=|\Omega_{p0}|^2\frac{\beta^2}{\kappa^2}[\sinh{(\kappa z)}]^2e^{-\eta z},\\
|\Omega_s(z)|^2=|\Omega_{p0}|^2[\cosh{(\kappa z)}+\frac{\eta}{2\kappa}\sinh(\kappa z)]^2e^{-\eta z},    
\end{eqnarray}
where $\eta=\frac{\alpha}{2L}$ and $\kappa=\sqrt{(\frac{\eta}{2})^2-\beta^2}$. 
If $\eta\gg\beta$ (i.e., the OD is sufficiently large), these solutions at the trailing end of medium can be reduced to
\begin{eqnarray}
|\Omega_p(z=L)|^2=|\Omega_{p0}|^2\frac{\pi^2}{\alpha^2}(1-e^{-\frac{\alpha}{2}})^2,\\
|\Omega_s(z=L)|^2=|\Omega_{p0}|^2(1-\frac{\pi^2}{\alpha}).
\end{eqnarray}
The theoretical curves of Eqs. (13) and (14) are plotted in Fig.~\ref{fig:Theoretical curves} to demonstrate the CE can reach 96\% at an OD of 240, which is close to that of the backward FWM scheme that employs an identical OD \cite{Liu17}.

\FigThree


\section{Intensity Mismatch} \label{Sec:Int}

To demonstrate the SLM-based FWM in the experiment, we can create a similar condition to that of Eqs. (11) and (12) by arranging the propagating direction of $\Omega_{c}$ and $\Omega_{d}$, as shown in Fig.~\ref{fig:Intensity mismatch}(a); the direction is perpendicular to that of $\Omega_p$. $\Omega_{c}$ and $\Omega_{d}$ irradiate at the leading edge and trailing edge of the medium, respectively. Produced by the SMF, the strengths of $\Omega_{c}$ and $\Omega_{d}$ in space would exhibit Gaussian distributions on the z-axis. Because an angle exists between $\Omega_p$ and the two control beams, $\Omega_p$ encounters a spatially varied condition of $\Omega_{c}$ and $\Omega_{d}$ if the waists of the two control beams are carefully selected and matched with the atomic medium length ($L$). However, more energy is lost in the FWM system due to the Doppler effect and phase-mismatch effect as the angle between the probe and the two control beams increases. Because of the random motions of cold atoms with a temperature of approximately 300 $\mu$K in our experiment, a non zero value of two-photon detuning in the EIT system is induced if $\Omega_p$ and $\Omega_c$ are not in parallel. In addition, if $\Omega_c$ and $\Omega_d$ are in parallel but perpendicular to $\Omega_p$, as shown in Fig.~\ref{fig:Intensity mismatch}(a), the phase-mismatch term is large and cannot be neglectged in the double-$\Lambda$ FWM system \cite{Liu17}. Overall, the two effects would substantially reduce the probe and signal transmissions in the SLM-based FWM scheme.

\FigFour

\FigFive

To prevent the aforementioned two unwanted effects, the angle between the probe and the two control beams should be as narrow as possible, and thus an angle of approximately 2$^\circ$ is chosen in our experiment. Based on our numerical simulation, the energy loss caused by these two effects drops to below 1\% when this narrow angle is employed in the current experimental conditions. Therefore, we have excluded these two effects from our theoretical model. 

For $L \approx 3.5$ mm in our MOT system, the required waists of $\Omega_c$ and $\Omega_d$ are calculated as approximately 120 $\mu$m when the angle is set to $2^\circ$. However, $\Omega_{c}$ and $\Omega_{d}$ in this condition generate another effect that lowers the CE in our experimental observations. We termed this the intensity-mismatch effect in the SLM-based FWM system. To illustrate, $\Omega_{c}$ and $\Omega_{d}$ cannot meet our desired intensity distribution for the entire spatial region of $\Omega_p$ when the angle is set to 2$^\circ$, as shown in Fig.~\ref{fig:Intensity mismatch}(b). In other words, the edge of $\Omega_p$ faces the shifted intensity modulation and thus has a lower CE relative to the peak region of $\Omega_p$. This phenomenon reduces the CE in our current experimental conditions compared with the ideal conditions. Nevertheless, in principle, the intensity-mismatch effect can be mitigated by using $\Omega_p$ with a smaller waist.


\section{Results and Discussion} \label{Sec:result}

We first measure the slow light pulses in the EIT experiment to determine some experimental parameters. Fig.~\ref{fig:Normal FWM}(a) shows the experimental data of the probe pulses propagating through the EIT medium. The black and red solid lines are the incident and transmitted probe pulses, respectively. The dashed lines are the theoretical curves with parameters of $\alpha = 19$, $\Omega_c = 0.26\Gamma$, and $\gamma_{21} = 5\times 10^{-4}\Gamma$. Under the same experimental circumstances, we activate the driving field under the condition of $\Omega_d \approx \Omega_c$ to perform the FWM experiment. Notably, the coupling and driving beams completely overlap ($\Delta$S = 0). Because the strengths of $\Omega_c$ and $\Omega_d$ are almost equal, we can obtain the maximum CE in the resonant FWM processes according to the theoretical prediction. The experimental observation is shown in Fig.~\ref{fig:Normal FWM}(b) and demonstrates that the transmissions of the probe and signal fields reach approximately 20\%, which is slightly lower than the conversion limit (25\%) because of the non negligible ground state dephasing rate. In addition, the dephasing rate increases from $5 \times 10^{-4}\Gamma$ to $1.0 \times 10^{-3}\Gamma$ when the driving field is activated, indicating that $\Omega_d$ slightly damages the EIT condition through the far-detuning photon-switching effect \cite{Harris98}. Figure~\ref{fig:Normal FWM}(b) also shows that the group velocity of the generated signal pulse is not identical to that of the transmitted probe pulse because of fluctuatuions of the control laser intensities and OD in each measurement.

\FigSix

To realize the SLM-based FWM, two CLs are applied to focus the waists of $\Omega_c$ and $\Omega_d$. As previously, we first conduct the slow light experiment to determine some experimental parameters, as shown in Fig.~\ref{fig:SLM FWM}(a). The experimental observation shows that the probe transmission clearly decreases compared with that in Fig.~\ref{fig:Normal FWM}(a) because $\Omega_p$ propagates through a relatively small $\Omega_c$ in both the leading and trailing parts of the atomic cloud with a non zero dephasing rate ($\gamma_{21}\neq 0$); this causes additional energy loss of approximately 15\% in the current experimental conditions. The theoretical curves for the experimental data shown in Fig.~\ref{fig:SLM FWM} have taken the 3D distributions of $\Omega_p$, $\Omega_c$, and $\Omega_d$ into the considerations.

By introducing $\Omega_{d}$ with the same beam profile and waist of $\Omega_{c}$, we first perform the SLM-based FWM experiment with $\Delta$S = 3 $\mu$m; in other words, two control beams are almost focused on the center of the atomic cloud. Subsequently, $\Omega_{c}$ and $\Omega_{d}$ are separated by adjusting the position of each CL. The experimental accuracy of $\Delta$S determined through a rotating knife-edge method is approximately 1 $\mu$m. In total, we measure five different $\Delta$S, namely 3, 30, 54, 75 and 95 $\mu$m; the experimental results with each value of $\Delta$S are plotted in Fig.~\ref{fig:SLM FWM}(b)-(f). The delay time of the slow probe and signal pulses increase as the two control beams are gradually separated.

The transmissions of the probe and signal pulses propagating in the SLM-based FWM medium versus $\Delta$S are plotted in Fig.~\ref{fig:SLM FWM Efficiency}. The red circles and blue squares represent the probe and signal transmissions, respectively. The error bar represents a $\pm$1 standard deviation based on five successive measurements. The solid lines are the theoretical curves with the following parameters: $\alpha=19$, $\Omega_{c}=0.39\Gamma$, $\Omega_{d}=0.41\Gamma$, and $\gamma_{21}=8 \times 10^{-4}\Gamma$. The dashed lines indicate theoretical curves with the $\pm$1 standard deviation of $\gamma_{21}$ [i.e., $\gamma_{21}=5 \times 10^{-4}\Gamma$ (upper line) and $\gamma_{21}=1.1 \times 10^{-3}\Gamma$ (lower line)]. We observe a maximum CE of 43 $\pm$ 2\% when $\Delta$S = 54 $\mu$m, which is in strong agreement with the 3D theoretical simulation, as shown in Fig.~\ref{fig:SLM FWM Efficiency}. Additionally, because the modulated intensities of two control beams exhibit the Gaussian distributions along the z-axis, the maximum CE is not identical to that obtained by the theoretical model discussed in Section~\ref{Sec:thy} which uses sine and cosine functions.

\FigSeven


\section{CONCLUSION} \label{Sec:conclusion}

This study demonstrates wavelength conversion from 780 to 795 nm with a 43\% CE at an OD of 19 in cold $^{87}$Rb atoms by employing SLM-based FWM scheme. This scheme can overcome the conversion limit in the resonant FWM processes. Furthermore, according to the theoretical model, a 96\% CE can be realized at an OD of 240 under ideal conditions. Such a highly efficient FWM scheme can achieve a near-unity CE, thereby providing an easy method of implementing a high-fidelity quantum wavelength converter for narrowband single photons in all-optical quantum information processing.


\section*{ACKNOWLEDGEMENTS}
We thank Ying-Cheng Chen, Dian-Jiun Han, and Chung-Hsien Chou for their useful contributions to discussions. This work was supported by the Ministry of Science and Technology of Taiwan under Grants No. 106-2119-M-006-002. We also acknowledge the support from NCTS of Taiwan.


\end{document}